\documentclass[11pt,preprint2]{aastex}

\shorttitle{\ion{Fe}{2}/\ion{Mg}{2} of QSOs}
\shortauthors{Iwamuro et al.}
\begin{document}
\title{\ion{Fe}{2}/\ion{Mg}{2} EMISSION LINE RATIOS OF QSOs. II. $z>6$ OBJECTS}
\author{\sc Fumihide Iwamuro\altaffilmark{1}, Masahiko Kimura\altaffilmark{1}, Shigeru Eto\altaffilmark{1}, Toshinori Maihara\altaffilmark{1}, 
Kentaro Motohara\altaffilmark{2}, Yuzuru Yoshii\altaffilmark{2,3}, and Mamoru Doi\altaffilmark{2}}
\altaffiltext{1}{Department of Astronomy, Kyoto University, Kitashirakawa, Kyoto 606-8502, Japan}
\altaffiltext{2}{Institute of Astronomy, School of Science, The University of Tokyo, 2-21-1 Osawa, Mitaka, Tokyo 181-0015, Japan}
\altaffiltext{3}{Research Center for the Early Universe, School of Science, University of Tokyo, Tokyo 113-0033, Japan}
\begin{abstract}
Near-infrared spectra of four QSOs located at $z>6$ are obtained with the OH-airglow suppressor mounted on the Subaru telescope. 
The \ion{Fe}{2}/\ion{Mg}{2} emission-line ratios of these QSOs are examined by the same fitting algorithm as in our previous 
study of $z<5.3$ QSOs. The fitting results show that two out of the four $z>6$ QSOs have significant \ion{Fe}{2} emission in their 
rest-UV spectra, while the other two have almost no \ion{Fe}{2} features. We also applied our fitting algorithm to more than 10,000 SDSS QSOs 
and found two trends in the distribution of \ion{Fe}{2}/\ion{Mg}{2} against redshift: (1) the upper envelope of the \ion{Fe}{2}/\ion{Mg}{2}
distribution at $z>3$ shows a probable declination toward high redshift, and (2) the median distribution settles into lower ratios at $z\sim 1.5$ 
with small scatter compared to the other redshift. We discuss an Fe/Mg abundance evolution of QSOs with a substantial contribution 
from the diverse nature of the broad-line regions in high-redshift QSOs.
\end{abstract}

\section{INTRODUCTION}
The Fe/Mg abundance ratio, one of the most important parameters for estimating the age of a stellar system after its initial starburst, 
is based on the delay of iron enrichment by Type Ia supernovae and compared with the formation of such $\alpha$-elements as magnesium 
by Type II supernovae (see Hamann \& Ferland 1999 for a review). The delay of iron formation was generally thought to be $\sim 1$ Gyr 
(e.g., Yoshii et al. 1996), while more recent studies suggest that this delay can be $\sim 0.3$ Gyr, depending on environmental 
conditions of the star formation process \citep{fri98,mat01}. The \ion{Fe}{2}(UV)/\ion{Mg}{2} emission-line ratio of QSOs is considered 
a probable indicator of the Fe/Mg abundance ratio, which has been measured in various redshift ranges: $3.1<z<4.7$ by \citet[hereafter THE99]{tho99}, 
$0<z<5.3$ by \citet[hereafter Paper~I]{iwa02}, $5.7<z<6.3$ by \citet{fre03}, $0<z<4.8$ by \citet{die03}, and $3.0<z<6.4$ by \citet{mai03}.
The median values of the \ion{Fe}{2}/\ion{Mg}{2} emission-line ratios are almost constant at all redshifts, while large scatter dominates 
the distribution of the ratios at $z>3$, indicating the diversity of the formation histories of high-redshift QSOs. On the other hand, recent 
photoionization calculations of Fe$^+$ atoms in the broad-line regions of QSOs show that the \ion{Fe}{2}(UV)/\ion{Mg}{2} emission-line 
ratio is sensitive not only to Fe and Mg abundance but also to physical conditions, including microturbulence \citep{ver03}. However, 
the median value of the \ion{Fe}{2}/\ion{Mg}{2} emission-line ratio is expected to converge on a significantly smaller value when the age of 
the universe becomes less than $\sim 0.3$ -- 1 Gyr, assuming that the substantial scatter caused by the diverse characteristics (or microturbulence 
velocities, etc.) of the broad-line regions in QSOs has the same contribution at any redshift.

In this paper, we report the \ion{Fe}{2}/\ion{Mg}{2} emission-line ratios of four QSOs at $z>6$ observed by the OH-airglow suppressor 
(OHS; Iwamuro et al.\ 2001) and CISCO \citep{mot02} mounted on the Subaru telescope. In $\S2$ we analyze the data by the same method 
as in Paper~I, and then we compare the results with other observations in $\S3$. In $\S4$ we discuss the expected Fe/Mg abundance 
evolution of QSOs using the combined samples of these results, the new Sloan Digital Sky Survey (SDSS) archival data\footnote{See http://www.sdss.org/dr1.}, 
and the previous samples in Paper~I. Throughout the paper we adopt a cosmology with $H_0=70$ km s$^{-1}$ Mpc$^{-1}$, $\Omega_M=0.3$, and 
$\Omega_{\Lambda}=0.7$.

\section{OBSERVATIONS AND DATA REDUCTION}
The observations were carried out on 2002 February~27 and March~2 and again on 2003 March~20 and 21, using OHS and CISCO mounted on the Subaru telescope.
Since the \ion{Fe}{2} and \ion{Mg}{2} are redshifted to the $H$ and $K$ bands, respectively, for $z>6$ QSOs, the $K$-band spectra were 
obtained by CISCO and the $JH$-band spectra by OHS separately in each observation run. The typical exposure sequence is 200~s $\times$ 3 
$\times$ 4 positions (CISCO) or 1000~s $\times$ 4 positions (OHS), in which the object is moved about $10\arcsec$ along the slit by 
nodding the telescope after every three exposures (CISCO) or one exposure (OHS). The slit widths were $0\farcs 8$ (CISCO) and $1\arcsec$ (OHS), 
providing spectral resolutions of 330 and 210, respectively. After this exposure sequence, a nearby SAO star with a spectral type 
of A or F was observed as a spectroscopic standard to remove the telluric atmosphere absorption features and to correct the instrumental 
response. The pixel scale was $0\farcs 106$/pixel with the infrared secondary mirror of the telescope, while the seeing size was 
$0\farcs 7$. To calibrate the relative flux between the $K$- and $JH$-band spectra, short imaging exposures of 20~s $\times$ 3 
$\times$ 4 positions were executed by CISCO in both the $K'$ (1.97--2.30~$\mu$m) and $H$ (1.50--1.79~$\mu$m) bands, except for SDSS~1030+0524. 
The observation log is summarized in Table~1.

The obtained data were reduced using IRAF with typical reduction procedures for infrared images (see Paper~I). The resultant one-dimensional 
spectra in the $K$ and $JH$ bands were combined using the photometric results of the imaging observation by CISCO. As a result, through a 
circular aperture of 2\farcs2 diameter, the average fluxes between 1.97--2.30 and 1.50--1.79~$\mu$m correspond to the $K'$- and $H$-band 
magnitudes respectively. The extraction procedure for \ion{Fe}{2} and \ion{Mg}{2} emission lines from the combined spectra is the same as 
in Paper~I: nonlinear $\chi^2$ fitting for the \ion{Mg}{2} (Moffat function) and for the continuum (power law) profile with six free parameters, 
after subtraction of the \ion{Fe}{2} template from the composite QSO spectrum of the Large Bright Quasar Survey (LBQS; Francis et al. 1991) and the Balmer 
continuum template from the UV spectrum of 3C~273 \citep{wil85} with various multiplying factors (see eq. (1)--(4) in Paper~I). The fitting 
range is 2150--3300\AA, which is also the wavelength range for our definition of \ion{Fe}{2} flux. Note that the strength of the \ion{Fe}{2} 
emission is determined by shape comparison between our \ion{Fe}{2} template and the observed spectral features (large bumps and small 
depressions), which does not require the assumption of the power-law continuum level (we need the assumption of the shape of the \ion{Fe}{2} 
template instead).

Although the Moffat function is very useful for fitting the various profiles of \ion{Mg}{2} emission lines with the minimum number of parameters, 
a serious problem is that the fitted \ion{Mg}{2} profile sometimes has very broad wings. To prevent such unexpected broad wings, 
we set the lower limits at $b\geq 0.5$ for the Moffat power (eq. (1) in Paper~I) and reanalyzed all the data fitted with $b<0.5$ in 
Paper~I (see Appendix).

\section{RESULTS}
The object spectra with fitted components are shown in Figure~1, and the numerical results are listed in Table~2. The spectra with 
wavelengths longer than 1.80~$\mu$m are obtained by CISCO, and the shorter parts by OHS. The strength of the \ion{Fe}{2} emission is 
mainly determined by the break feature at rest 2200\AA, which is observed by OHS with superior signal-to-noise ratios. The wavelength range between 
1.80 and 1.95~$\mu$m corresponds to the atmospheric absorption band, whose contribution to the fitting results is very small because of 
large errors. 

As shown in Figure~1 and Table~2, we detected strong \ion{Fe}{2} emission from SDSS~J1148+5251, which is consistent with \citet{mai03}. 
SDSS~J1048+4637 also shows significant \ion{Fe}{2} emission, while the \ion{Fe}{2}/\ion{Mg}{2} ratio is not as much as the value of $\sim$8.1 
reported by \citet{mai03}. The absorption-line feature at 1.29~$\mu$m in SDSS~J1048+4637 is the broad absorption line component of 
\ion{C}{3}] reported by \citet{mai04}. The observed spectra of these objects fit well into our template spectrum of \ion{Fe}{2} in every detail.
Although \citet{bar03} reported a smaller \ion{Fe}{2}/\ion{Mg}{2} ratio of $\sim$4.7 for SDSS~J1148+5251, this inconsistency is mostly caused by 
the difference in the \ion{Fe}{2} template under \ion{Mg}{2} emission. On the other hand, SDSS~J1030+0524 and SDSS~J1630+4012 show 
almost no \ion{Fe}{2} emission or any break feature at rest 2200\AA. These results for SDSS~J1030+0524 are consistent with Freudling et 
al.(2003), as well as the unknown absorption-line feature at 1.57~$\mu$m. As a result, even at $z>6$, the \ion{Fe}{2}(UV)/\ion{Mg}{2} emission-line 
ratios still show large diversity.

\section{DISCUSSION}
Figure~2 shows the \ion{Fe}{2}/\ion{Mg}{2} ratio and the absolute AB magnitude at rest 2500\AA\ as a function of redshift, for which we 
reanalyzed the SDSS data on the basis of the DR1 quasar catalog \citep{sch03}. The other $z<5.3$ data are the same as those plotted in 
Figure~7 of Paper~I, except for four objects listed in the Appendix. The median values and the standard deviation of the sample distribution 
are shown by the squares with the cross bars in this figure whose numerical values are listed in Table~3. Although the number of samples 
at $z>3$ is insufficient to make a statistical argument, we can deduce the trends of the evolution of the \ion{Fe}{2}/\ion{Mg}{2} ratio 
from Figure~2. First, the median \ion{Fe}{2}/\ion{Mg}{2} ratios at $z>3$ are almost constant, while the upper envelope of the distribution 
plotted by the dashed line in Figure~2 shows a probable declination toward high redshift. Second, the median distribution settles into 
lower ratios at $z\sim 1.5$, with small scatter compared to the other redshift. Here, the downward error bars from these median points 
are almost equal to the typical fitting errors of the samples in each bin (column 3 in Table~3), while the longer upward error bars are 
affected by the substantial scatter of the \ion{Fe}{2}/\ion{Mg}{2} ratios.

Assuming that the Fe/Mg abundance ratio affects these rough trends, these small \ion{Fe}{2}/\ion{Mg}{2} ratios at $z\sim 1.5$ originated 
with the dilution of the Fe abundance with the outflow gas from low-mass stars after $\sim$3 Gyr from the initial starburst \citep{yos98}.
The scatter of the sample at this redshift (a factor of $\sim$2 between the top and bottom of the cross bars) corresponds to the maximum 
contribution of the scatter caused by the differences in the physical conditions of the broad-line regions in high-redshift QSOs 
\citep{ver03}. If this substantial scatter of factor of $\sim$2 is the universal nature of the QSOs, independent of the redshift, the 
larger scatter of the $z>3$ samples are affected by the difference in the Fe/Mg abundance ratio or the time passage after the initial starburst.
This idea is also consistent with the declination of the upper envelope of the \ion{Fe}{2}/\ion{Mg}{2} distribution toward high redshift, 
which is expected to be the initial abundance evolution of QSOs.

An alternative explanation for the variation of the \ion{Fe}{2}/\ion{Mg}{2} ratios at $z\sim 2$ may be their luminosity dependence such that 
the QSOs of higher luminosities, in a low-redshift ($z=$0.1-0.6) sample from {\it Hubble Space Telescope} ($HST$) UV archives \citep{tsu04} 
as well as in a high-redshift ($z=$3-5) 
sample from near-infrared observations \citep{iwa02,die03}, are found to have higher \ion{Fe}{2}/\ion{Mg}{2} ratios. The larger variation of the 
\ion{Fe}{2}/\ion{Mg}{2} ratios for the brighter QSOs suggests that either the ionizing photon flux arriving at the broad-line-emitting clouds 
or the physical characteristics of the clouds would scatter more significantly for brighter QSOs. To estimate their respective contributions 
to the \ion{Fe}{2}/\ion{Mg}{2} ratio is a theoretical challenge that requires an elaborate photoionization model of Fe$^+$ in the broad-line 
regions of QSOs.

\acknowledgments
We thank the Canadian Astronomy Data Centre, which is operated by the Herzberg Institute of Astrophysics, National Research Council of Canada. 
We would like to express our thanks to the members of the SDSS project. 

Funding for the SDSS has been provided by the Alfred P. Sloan Foundation, the Participating Institutions, 
the National Aeronautics and Space Administration, the National Science Foundation, the US Department of Energy, the Japanese Monbukagakusho, 
and the Max Planck Society. The SDSS Web site is http://www.sdss.org.

The SDSS is managed by the Astrophysical Research Consortium (ARC) for the Participating Institutions. The Participating Institutions are the 
University of Chicago, Fermilab, the Institute for Advanced Study, the Japan Participation Group, the Johns Hopkins University, Los Alamos 
National Laboratory, the Max-Planck-Institute for Astronomy (MPIA), the Max-Planck-Institute for Astrophysics (MPA), New Mexico State University,
the University of Pittsburgh, Princeton University, the United States Naval Observatory, and the University of Washington.

This work has been supported by a Grant-in-Aid for Scientific Research (A), Japan (14204016).
This work is also supported by a Grant-in-Aid for the 21st Century COE ``Center for Diversity and Universality in Physics,''
and in part by a Grant-in-Aid for Center-of-Excellence Research (07CE2002) of the Ministry of Education, Culture, Sports, Science, and Technology 
(MEXT) of Japan.

\appendix
\section{Re-analyzed Objects Having Broad \ion{Mg}{2} Wings}
We re-analyzed all the data having broad \ion{Mg}{2} wings ($b<0.5$) in Paper~I with lower limits at $b\geq 0.5$ for the Moffat profile of
\begin{equation}
F_{MgII}(\lambda)=i\left[1+\left(\frac{\lambda - 2798\AA}{a}\right)^2\right]^{-b}.
\end{equation}
All the data were fitted with $b=0.5$ (Lorentzian), and the results are listed in Table~4.
Three out of seven objects listed in Table~4 are not plotted in Figure~2 because we adopted the results of the OHS data rather than those of THE99 and 
the results of the $HST$ FOS archive rather than those of \citet{kin91} (see Table~2 and Table~4 in Paper~I).

\clearpage
\onecolumn
\begin{figure}
\begin{minipage}{18cm}
\epsscale{0.9}
\plottwo{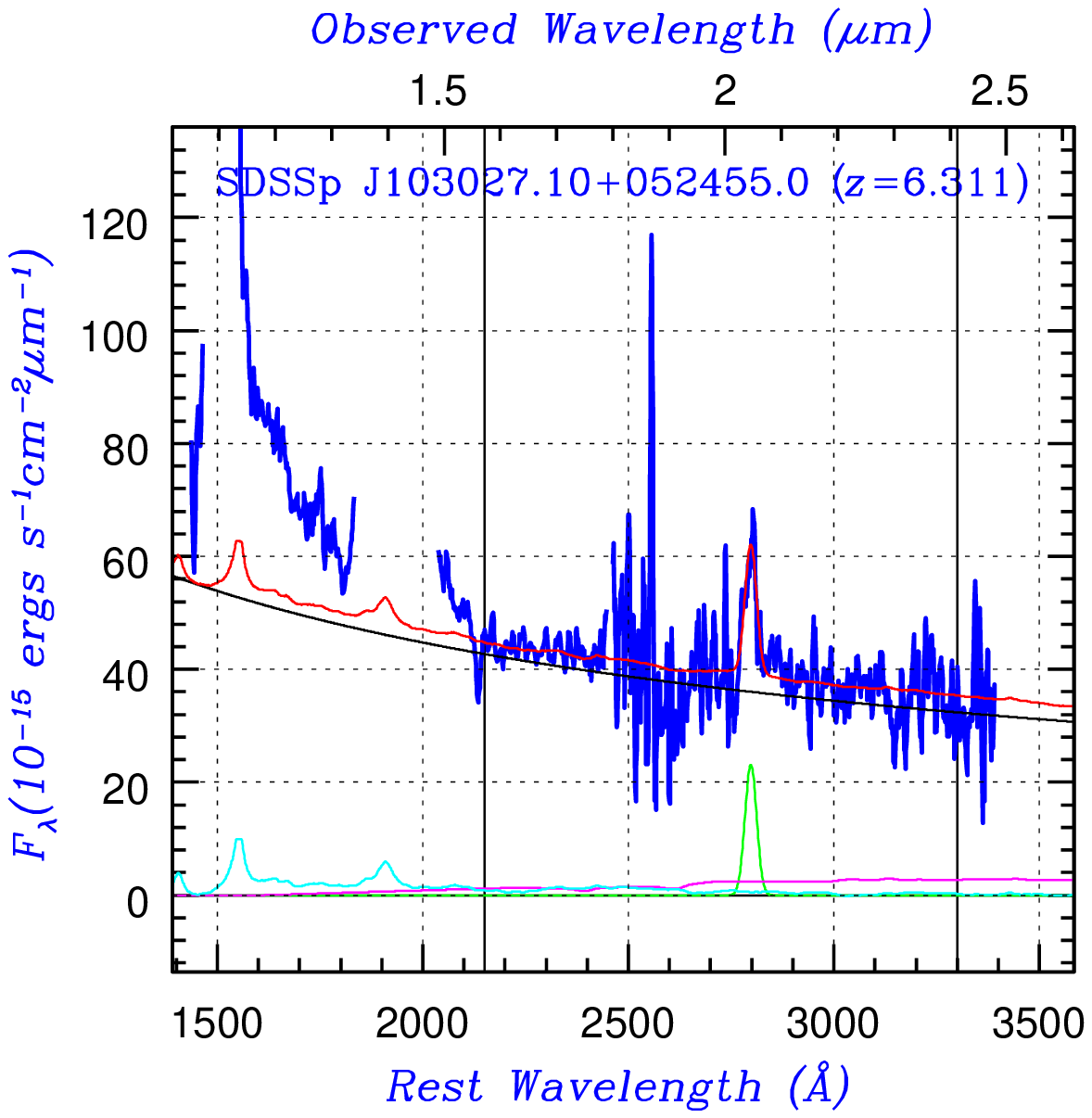}{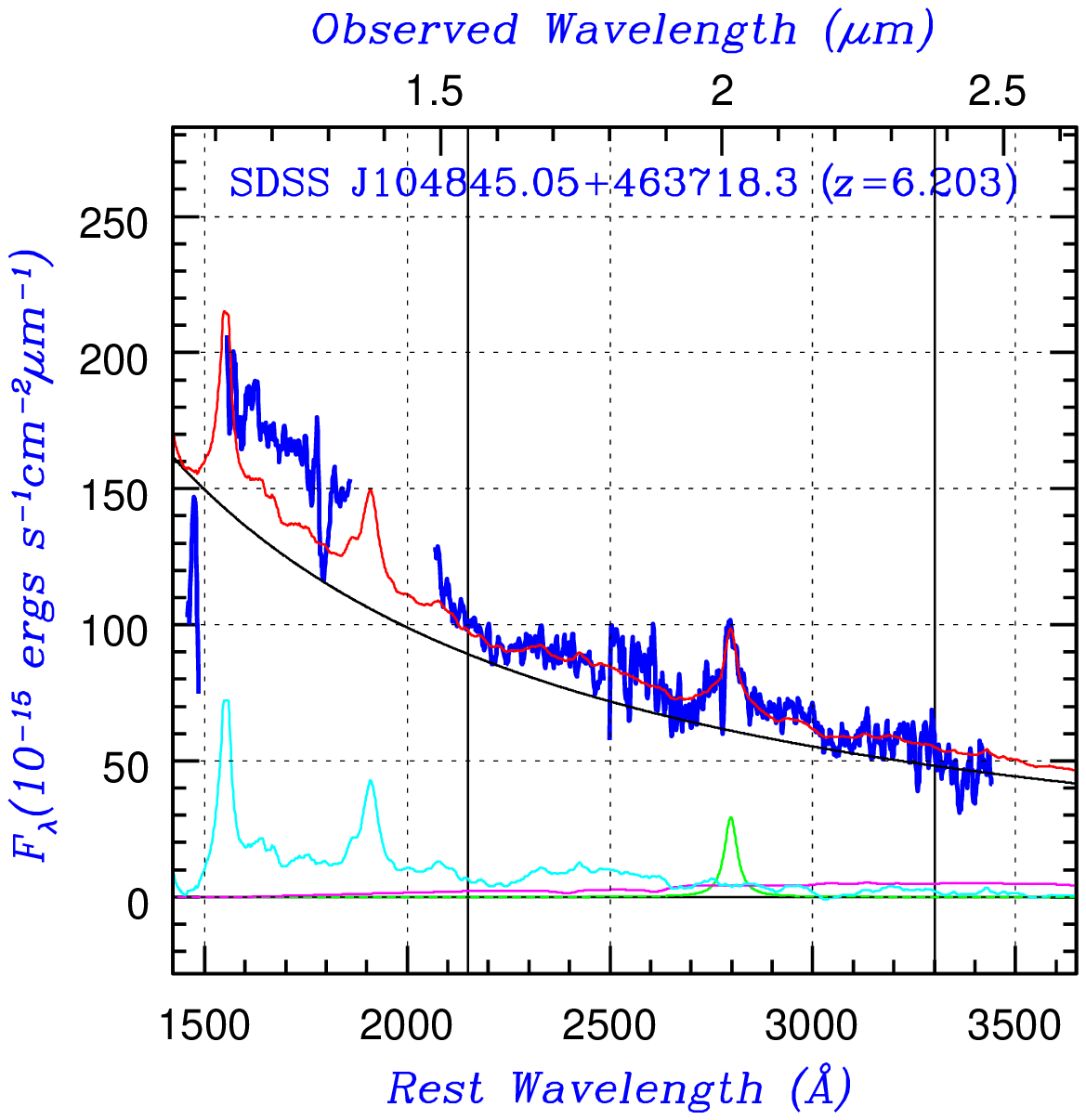}
\end{minipage}
\begin{minipage}{18cm}
\epsscale{0.9}
\plottwo{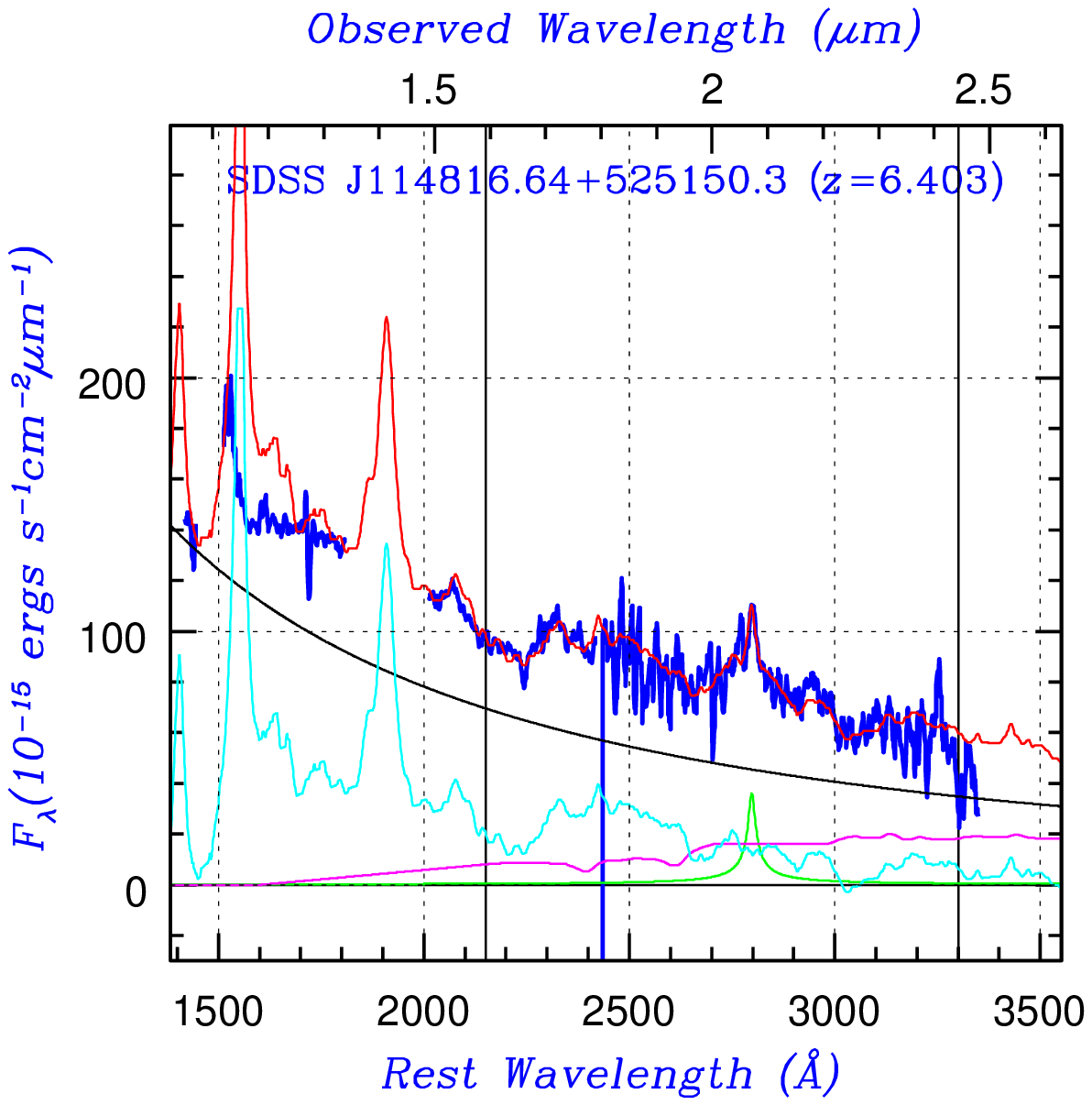}{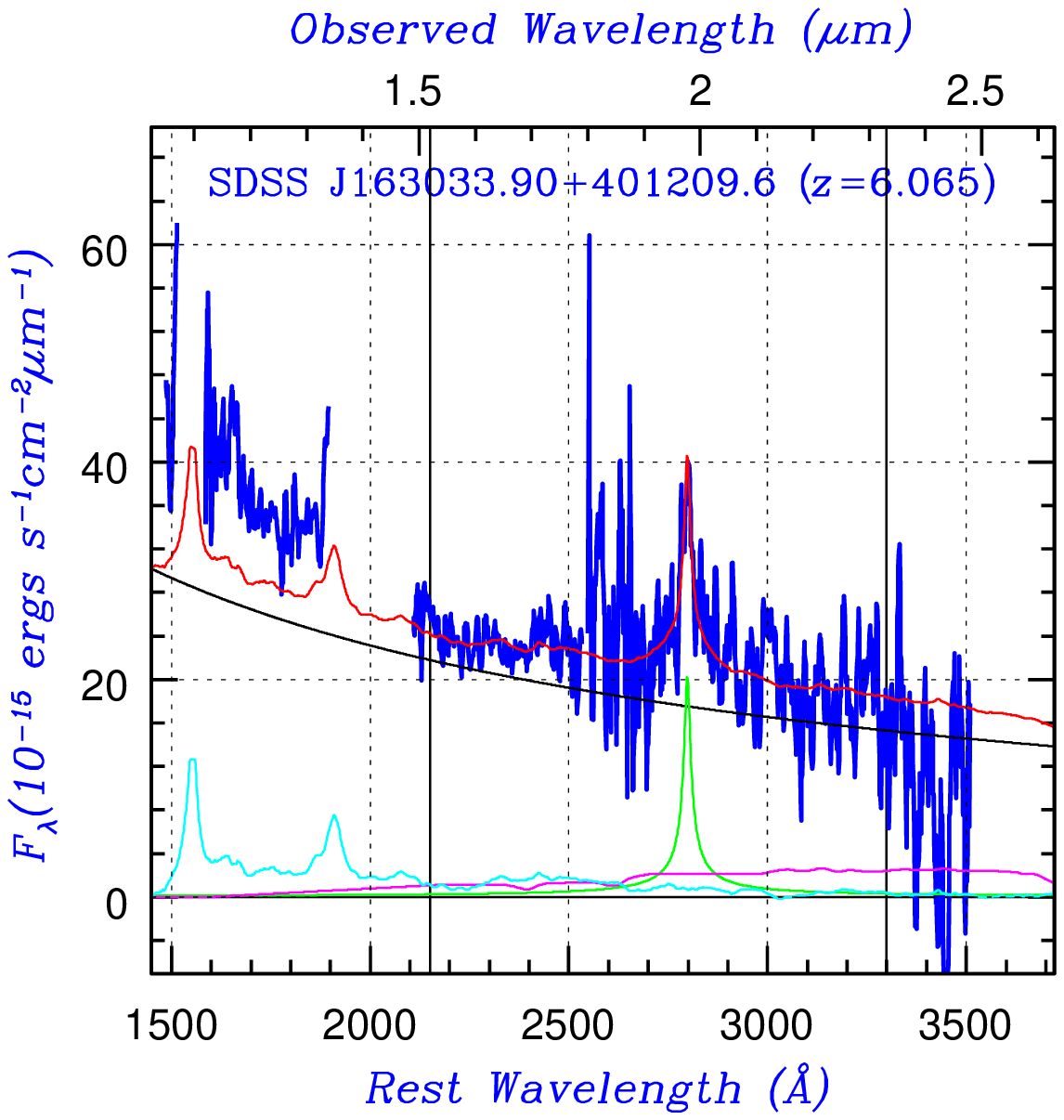}
\end{minipage}
\caption{Observed spectra of the four QSOs listed in Table 1. 
A power law continuum, \ion{Mg}{2}, \ion{Fe}{2}, a Balmer continuum (including other lines except for \ion{Mg}{2}),
and the sum of these components are plotted for comparison with the observed spectrum with the boxcar-smoothing 
of six pixels. The two vertical lines indicate the fitting range of 2150--3300\AA\ in the rest-wavelength.}
\end{figure}
\clearpage
\begin{figure}
\plotone{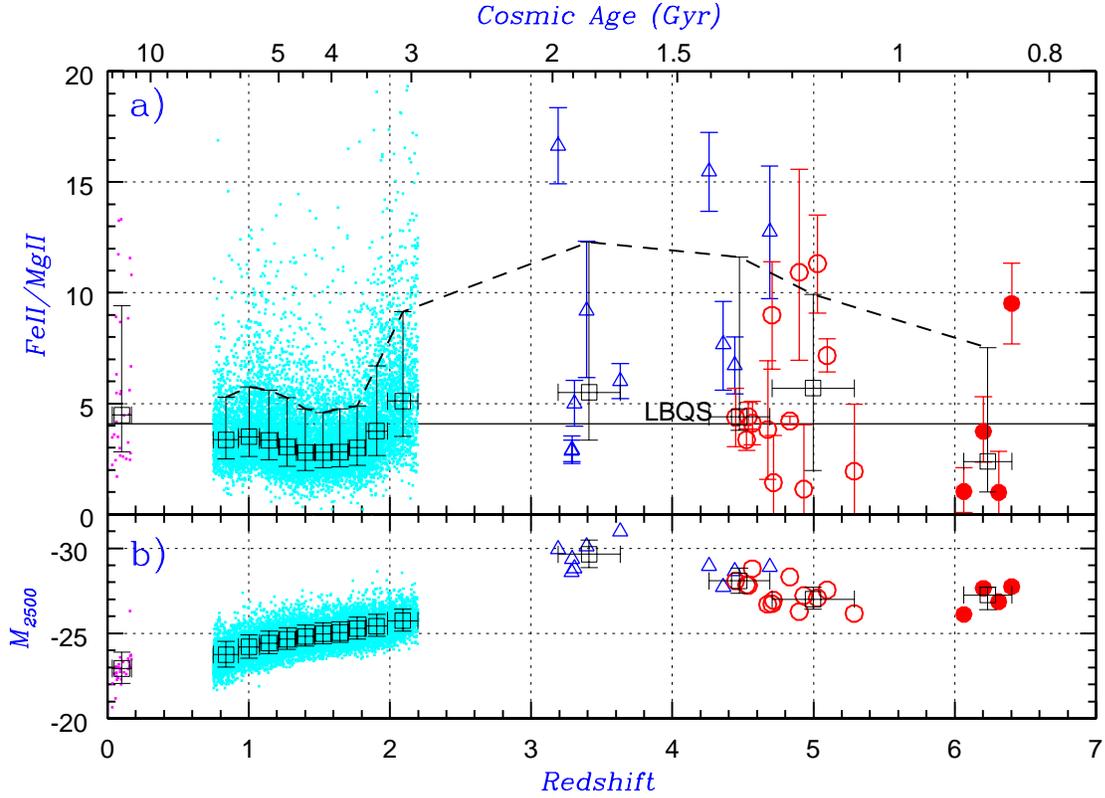}
\caption{Redshifts vs. \ion{Fe}{2}/\ion{Mg}{2} and absolute magnitudes. The fitting results for the QSO spectra from \citet{kin91}, 
$HST$/FOS and SDSS archives (dots), THE99 (triangles), our data in Paper~I (open circles), and this work 
(filled circles) are plotted with the median values (squares) of the appropriate bins. The horizontal solid line indicates the 
\ion{Fe}{2}/\ion{Mg}{2} of 4.09 for the LBQS composite spectrum (see Paper~I). The squares with the cross-bars represent the median 
values and the standard deviations of the sample distribution, whose upper envelope is connected by a broken line.
We assume $H_0=70$ km s$^{-1}$ Mpc$^{-1}$, $\Omega_M=0.3$, and $\Omega_{\Lambda}=0.7$ for the estimation of $M_{2500}$ (the absolute 
AB-magnitude at rest-frame 2500\AA) and the cosmic age at the redshift.}
\end{figure}

\clearpage
\begin{deluxetable}{lcccccccc}
\rotate
\tabletypesize{\small}
\tablecaption{\sc Log of Observations}
\tablewidth{0pt}
\tablehead{\colhead{} & \colhead{} & \colhead{} & \colhead{} & \colhead{} & \colhead{} & \colhead{Exposure} & \colhead{} & \colhead{Coordinate} \\
\colhead{Object Name} & \colhead{Redshift} & \colhead{$K'$-mag\tablenotemark{a}} & \colhead{$H$-mag\tablenotemark{a}} & \colhead{Date} & 
\colhead{Instrument\tablenotemark{b}} & \colhead{Time (s)} & \colhead{Seeing} & \colhead{Reference}}
\startdata
\objectname[]{SDSS~J103027.10+052455.0} & 6.28 & 17.67 &  ---\tablenotemark{c} & 2002 Feb.27 & CISCO & 2400 & 0\farcs80 & 1 \\
                                        &      &  ---  & 18.57 & 2002 Mar. 2 &  OHS  & 4000 & 0\farcs88 &\\
\objectname[]{SDSS~J104845.05+463718.3} & 6.23 & 17.12 & 17.76 & 2003 Mar.20 & CISCO & 2400 & 0\farcs56 & 2 \\
                                        &      &  ---  & 17.83 & 2003 Mar.21 &  OHS  & 4000 & 0\farcs71 &\\
\objectname[]{SDSS~J114816.64+525150.3} & 6.43 & 16.98 & 17.70 & 2003 Mar.20 & CISCO & 2400 & 0\farcs65 & 2 \\
                                        &      &  ---  & 17.62 & 2003 Mar.21 &  OHS  & 4000 & 0\farcs62 &\\
\objectname[]{SDSS~J163033.90+401209.6} & 6.05 & 18.40 & 19.25 & 2003 Mar.20 & CISCO & 2400 & 0\farcs71 & 2 \\
                                        &      &  ---  & 19.18 & 2003 Mar.21 &  OHS  & 8000 & 0\farcs68 &\\
\enddata
\tablenotetext{a}{Observed magnitudes by CISCO and OHS short-imaging observations with a 2\farcs2 circular diameter aperture. Typical photometric errors are 0.05 mag.}
\tablenotetext{b}{The slit widths are $0\farcs 8$ (CISCO) and $1\arcsec$ (OHS), corresponding to spectral resolutions of 330 and 210, respectively.}
\tablenotetext{c}{The $H$-band imaging observation was not executed.}
\tablerefs{(1) \citet{fan01}; (2) \citet{fan03}.}
\end{deluxetable}
\clearpage
\begin{deluxetable}{lccccccc}
\rotate
\tablecaption{\sc Results of Fitting Calculations}
\tablewidth{0pt}
\tablehead{\colhead{} & \colhead{} & \colhead{FWHM\tablenotemark{a}} & \colhead{$I$(\ion{Fe}{2})\tablenotemark{b}} & 
\colhead{$I$(\ion{Mg}{2})} && \colhead{$EW$(\ion{Fe}{2})\tablenotemark{b}} & \colhead{$EW$(\ion{Mg}{2})}\\
\colhead{Object Name} & \colhead{$z$} & \colhead{(km s$^{-1}$)} & \multicolumn{2}{c}{(10$^{-16}$ergs s$^{-1}$cm$^{-2}$)} & 
\colhead{\ion{Fe}{2}/\ion{Mg}{2}\tablenotemark{b}} & \colhead{(\AA)} & \colhead{(\AA)}}
\startdata
\objectname[]{SDSS~J103027.10+052455.0} & 6.311 & 3590 & 5.96$^{+11.1}_{-5.96}$ & 6.04$\pm$1.40 & 0.99$^{+1.86}_{-0.99}$ & 21.0 & 22.9\\
\objectname[]{SDSS~J104845.05+463718.3} & 6.203 & 4050 & 42.4$\pm$14.7          & 11.3$\pm$2.1  & 3.74$\pm$1.47          & 81.8 & 25.7\\
\objectname[]{SDSS~J114816.64+525150.3} & 6.403 & 3020 &  137$\pm$16            & 14.4$\pm$2.2  & 9.52$\pm$1.82          & 333  & 41.7\\
\objectname[]{SDSS~J163033.90+401209.6} & 6.065 & 2690 & 7.30$\pm$7.00          & 7.12$\pm$1.61 & 1.02$\pm$1.01          & 52.2 & 55.8\\
\enddata
\tablenotetext{a}{FWHM of the \ion{Mg}{2} emission-line in the rest-wavelength.}
\tablenotetext{b}{\ion{Fe}{2} is defined over the domain of 2150--3300\AA.}
\end{deluxetable}

\clearpage
\begin{deluxetable}{cccc}
\tablecaption{\sc Median values of {\rm \ion{Fe}{2}/\ion{Mg}{2}}}
\tablewidth{0pt}
\tablehead{\colhead{Redshift\tablenotemark{a}} & \colhead{\ion{Fe}{2}/\ion{Mg}{2}\tablenotemark{b}}
 & \colhead{Error\tablenotemark{c}} & \colhead{Number\tablenotemark{d}}}
\startdata
0.101$\pm$0.068 &  4.48$^{+4.95}_{-1.66}$ & 0.84 & 30\\
0.839$\pm$0.089 &  3.36$^{+1.93}_{-0.85}$ & 0.82 & 1101\\
1.003$\pm$0.075 &  3.49$^{+2.26}_{-0.89}$ & 0.90 & 1101\\
1.144$\pm$0.066 &  3.34$^{+2.25}_{-0.88}$ & 0.90 & 1101\\
1.273$\pm$0.063 &  3.05$^{+2.22}_{-0.89}$ & 0.92 & 1101\\
1.402$\pm$0.066 &  2.78$^{+1.98}_{-0.81}$ & 0.93 & 1101\\
1.528$\pm$0.060 &  2.79$^{+1.80}_{-0.70}$ & 0.81 & 1101\\
1.647$\pm$0.060 &  2.81$^{+1.94}_{-0.67}$ & 0.84 & 1101\\
1.769$\pm$0.063 &  3.00$^{+1.89}_{-0.81}$ & 0.95 & 1101\\
1.907$\pm$0.075 &  3.75$^{+2.95}_{-1.10}$ & 1.24 & 1101\\
2.091$\pm$0.109 &  5.11$^{+4.03}_{-1.60}$ & 1.67 & 1099\\
3.411$\pm$0.220 &  5.50$^{+6.77}_{-2.16}$ & 1.45 & 6\\
4.475$\pm$0.215 &  4.40$^{+7.22}_{-0.61}$ & 1.76 & 9\\
4.997$\pm$0.291 &  5.69$^{+4.25}_{-3.71}$ & 2.70 & 8\\
6.234$\pm$0.169 &  2.38$^{+5.14}_{-1.37}$ & 2.02 & 4\\
\enddata
\tablenotetext{a}{Redshift range of each bin.}
\tablenotetext{b}{Median value with standard deviation of the sample distribution.}
\tablenotetext{c}{Typical (root mean square) fitting errors of \ion{Fe}{2}/\ion{Mg}{2} ratios of the samples included in each bin.}
\tablenotetext{d}{Number of samples.}
\end{deluxetable}

\clearpage
\begin{deluxetable}{lcccccc}
\tablecaption{\sc Results of Fitting Calculations for THE99 data}
\tablewidth{0pt}
\tablehead{\colhead{} & \colhead{} & \colhead{FWHM} && \colhead{$EW$(\ion{Fe}{2})} & \colhead{$EW$(\ion{Mg}{2})} & \colhead{}\\
\colhead{Object Name} & \colhead{$z$} & \colhead{(km s$^{-1}$)} &\colhead{\ion{Fe}{2}/\ion{Mg}{2}} & 
\colhead{(\AA)} & \colhead{(\AA)} & \colhead{Reference}}
\startdata
\objectname[]{BR 1033$-$0327}  & 4.527 & 4050 & 2.34$\pm$0.78\tablenotemark{a} & 98.5 & 46.6 & THE99\\
\objectname[]{1358+391}        & 3.288 & 4610 & 2.86$\pm$0.49 & 110  & 46.1 & THE99\\
\objectname[]{PKS 2126$-$158}  & 3.290 & 3430 & 2.92$\pm$0.63 & 71.3 & 29.1 & THE99\\
\objectname[]{IIIZw 2}         & 0.089 & 3020 & 1.95$\pm$0.28\tablenotemark{a} & 155  & 79.1 & $IUE$\\
\objectname[]{PG 0844+349}     & 0.065 & 2320 & 6.31$\pm$0.23 & 238  & 40.4 & $HST$\\
\objectname[]{IRAS 1334+24}    & 0.109 & 3010 & 2.62$\pm$0.27 & 138  & 40.9 & $HST$\\
\objectname[]{MRC 2251-178}    & 0.064 & 4200 & 3.53$\pm$0.27\tablenotemark{a} & 424  & 143  & $IUE$\\
\enddata
\tablenotetext{a}{These data are not plotted in Figure~2 (see Appendix).}
\tablerefs{($IUE$) \citet{kin91} (ftp://dbc.nao.ac.jp/DBC/NASAADC/catalogs/3/3157/); ($HST$) $HST$ archive in CADC (http://cadcwww.dao.nrc.ca/hst/science.html); (THE99) Thompson et al.~(1999)}
\end{deluxetable}

\end{document}